\documentclass[preprint2]{aastex61}
\newcommand{\elsie}{\textit{Elsie}}
\newcommand{\celeste}{\textit{Celeste}}
\newcommand{\skara}{\textit{Skara Brae}}
\newcommand{\angkor}{\textit{Angkor}}
\newcommand{\KIC}{KIC~8462852}
\newcommand{\rp}{$r^{\prime}$}
\newcommand{\ip}{$i^{\prime}$}

\submitjournal{AJ}

\shorttitle{Dips of KIC 8462852}
\shortauthors{Bodman et al.}

\begin{document}

\title{The Variable Wavelength Dependence of the Dipping event of  KIC 8462852}


\correspondingauthor{Eva Bodman}
\email{ebodman1@asu.edu}

\author[0000-0002-4133-5216]{Eva H. L. Bodman}
\altaffiliation{NASA Postdoctoral Program Fellow, Nexus for Exoplanet System Science}
\affiliation{School of Earth and Space Exploration, Arizona State University, P.O. Box 871404, Tempe, AZ 85287-1404, USA}

\author{Jason Wright}
\affiliation{Department of Astronomy \& Astrophysics, The Pennsylvania State University, 525 Davey Lab, State College, PA 16802, USA}
\affiliation{Center for Exoplanets and Habitable Worlds, The Pennsylvania State University, 525 Davey Lab, University Park, PA 16802, USA}

\author[0000-0001-9879-9313]{Tabetha S. Boyajian}
\affiliation{Department of Physics and Astronomy, Louisiana State University, Baton Rouge, LA 70803 USA}

\author[0000-0001-6584-9919]{Tyler G. Ellis}
\affiliation{Department of Physics and Astronomy, Louisiana State University, Baton Rouge, LA 70803 USA}


\begin{abstract}
First observed with the \textit{Kepler} mission, KIC 8462852 undergoes unexplained dimming events, ``dips,'' on the timescale of days which were again observed from the ground from May to December 2017. Monitored with multi-band photometry by the Los Cumbres Observatory, all four dips of the ``\elsie\ dip family'' display clear wavelength dependence.  We measure how the wavelength dependence changes over the whole dimming event, including the dimming between the dips and the brightening event (the `blip') which occurs after the dips. We find that a single wavelength dependence does not fit the entire light curve and the dimming occurring between the dips is non-gray and varies in time. Because of the non-gray dimming between the dips, we measure the wavelength dependence of the dips separately and without the extra depth from this dimming. Such measurements yield a different estimate of the wavelength dependence the wavelength dependence of the dips but remains consistent with the previous measurement except for \elsie\ (the first dip), which is surrounded by dimming with strong wavelength dependence.   
We find the range of the wavelength dependence variation of the entire 2017 light curve is consistent with optically-thin dust with an average radius of $r<1\,\mu$m and the dust causing just the dips being $r<0.5\,\mu$m.
Since the dependence is time-dependent, the dust occulting the star must be heterogeneous in size, composition, or both and the distributions of these properties along the line of sight must change over time.
\end{abstract}

\keywords{stars: individual (KIC 8462852) --- stars: peculiar
}

\section{Introduction} \label{sec:intro}
An otherwise-typical F main sequence star, \KIC\ displayed a series of large decreases in observed flux (``dips''), up to 20\% during the \textit{Kepler} mission \citep{Borucki10}, that were discovered through the Planet Hunters citizen science project \citep{Boyajian2016}. The dips were asymmetric and the \textit{Kepler} light curve showed little evidence of periodicity in these events, although later observations and analysis showed hints \citep[e.g.][]{Simon2018, Kiefer2017, Bourne2018}. Archival and follow-up observations showed no evidence of infrared (IR) excess, putting strict limits on the amount of hot and warm dust in the system \citep{Boyajian2016, Marengo2015, Thompson2016}. From long cadence archival data from the Harvard plates, \citet{Schaefer2016} found that \KIC\ has dimmed $\sim$16\% over the last century, the so-called ``secular'' dimming and \citet{Castelaz2018} confirmed a slightly smaller secular dimming of $\sim12$\% with the archival plates from the Maria Mitchell Observatory. (However, \citet{Hippke2017} disputes the statistical significance of the \citet{Schaefer2016} results and finds no significant dimming in the Sonneberg plates).
Recent data also displays anomalous long-term and possibly periodic dimming behavior \citep{Meng2017,Simon2018} and \citep{Montet2016} analysis of the \textit{Kepler} full frame images show that the star dimmed by 3\% over the course of the mission.

To explain these large dips and sometimes the long term dimming, a myriad of causes were proposed and organized into general categories by \citet{Wright2016b}: circumstellar material, intervening material, stellar origins and solar system material. Circumstellar origins examined include large swarms of exocomets \citep{Bodman2016,Boyajian2016}, dust-enshrouded planetesimals on eccentric orbits \citep{Neslusan2017}, and a planetary ring system on a 12-year circular orbit with Trojans \citep{Ballesteros2018}. \cite{Foukal2017} studied the intrinsic variations. \citet{Katz2017} examined the plausibility of rings in the outer solar system. Few scenarios endeavor to plausibly explain both the long-term dimming and the short-term dips but \cite{Metzger2017} proposed the consumption of a secondary to naturally explain both anomalous behaviors. In this model, the star's luminosity would reduce after consuming the secondary and transiting material left over would result in dips. For general circumstellar material, \citet{Wyatt2018} showed that an elliptical dust ring with enough material to cause the secular dimming can be consistent with the IR constraints under certain conditions and clumps within such a ring can explain the dips with IR excess only detectable for about a week around the event. 

During a monitoring campaign funded by Kickstarter backers\footnote{https://www.kickstarter.com/projects/608159144/the-most-mysterious-star-in-the-galaxy}, \KIC\ began displaying dips from May to Oct 2017 which triggered a larger follow-up observation effort \citep{Boyajian2018}. This series of dips, dubbed the \elsie\ family, consists of four shallow dips of $\sim1-2\%$ in $r'$ band named \elsie, \celeste, \skara, and \angkor. Two more dips, \textit{Caral-Supe} and \textit{Evangeline}, occurred in 2018 that were deeper, up to $\sim5\%$, but in this work we restrict ourselves to the 2017 events. \citet{Boyajian2018} limited the chromatic analysis to the first dip \elsie\ and found the wavelength dependence of the transit is consistent with optically thin dust. \citet{Deeg2018} found a consistent wavelength dependence with their observations. The dust size required for the observed color is $<1\,\mu$m, which is smaller than the blow-out limit for an early F star, and therefore the dust must be newly created, if it  is circumstellar. The wavelength dependence for the dips is also larger than that observed for the long-term dimming \citep{Meng2017}, suggesting that these are two different dust populations.

Here, we perform a detailed analysis of the wavelength dependence of the \elsie\ family of dips with the LCOGT data (see section \ref{sec:observations}), restricting ourselves to the 2017 events. We study how the wavelength dependence of the dimming event changes with time in section \ref{sec:wholefit}. Then we examine each dip individually with the secular dimming removed in section \ref{sec:dips} and discuss in section \ref{sec:mix} how these measurements compare to those including the secular dimming. With the measured individual wavelength dependences of the dips, we estimate the grain size of the occulting dust assuming various compositions in section \ref{sec:grains}. Finally, we summarize our results.
%
%

\section{Observations} \label{sec:observations}
\begin{figure*}
\centering
\includegraphics[width=7.0in, trim= 0 0 0 0 ]{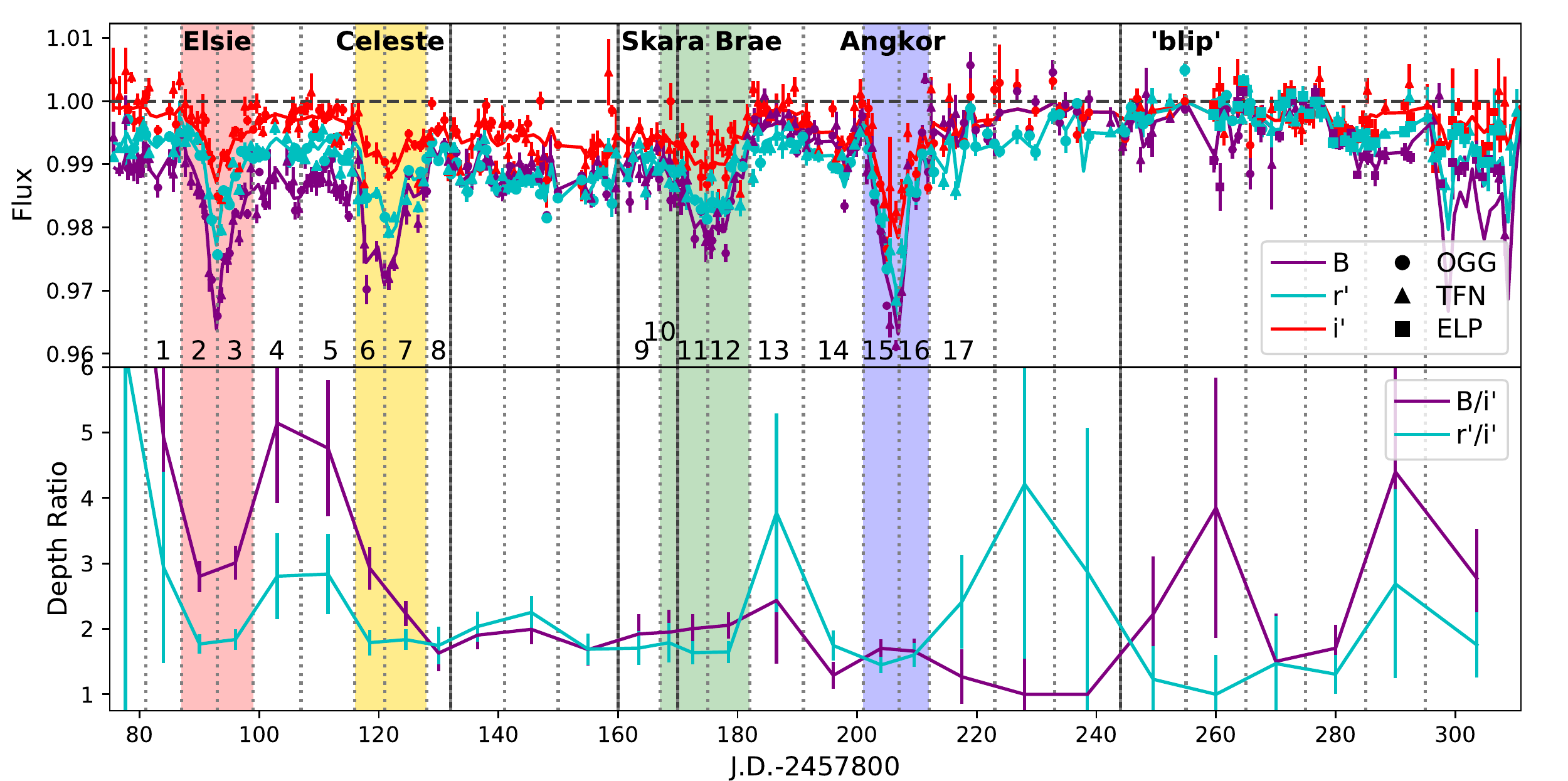}
\caption{The top plot is the light curve of the \elsie\ dip family with the fit described in detail in section \ref{sec:wholefit}. The three filters $B$, $r'$, and $i'$ are in purple, blue, and red, respectively. Each LCOGT site is marked with a different shape. The dotted vertical lines mark the bins for the depth ratios ($B/i'$ and $r'/i'$) and the solid, dark gray vertical lines mark the known discontinuities in the normalizations due to changes in instrumental setup. The \elsie, \celeste, \skara, and \angkor\ dips are highlighted in red, yellow, green, and blue, respectively. The label ``blip'' marks the approximate peak of the long term brightening trend occurring post-\angkor. The numbered regions refer to values listed in table \ref{tab:cols}. The bottom plot shows the depth ratios of each region over time with $B/i'$ in purple and $r'/i'$ in blue. Our measured depth ratios are insensitive to reasonable choices for the normalizations across the known discontinuities, and the fact that there is little or no evidence for discontinuities in the depth ratios across these boundaries demonstrates that our choices for normalizations are correct. A single depth ratio cannot fit the entire light curve well.
}
\label{fig:LC}
\end{figure*}

We acquired photometric time-series observations using the Las Cumbres Observatory (LCOGT) 0.4m robotic telescope network.  \citet{Boyajian2018} presented a subset of the observations described here, namely, data taken in $B$, \rp, \ip\ filters from 2017 May (during \elsie) as well as the time series in \rp\ from 2017 May to December when the object no longer was visible.  Here we provide a description of the full LCOGT 2017 data set taken during the \elsie\ family of dips. 
   
\KIC\ is visible from the northern hemisphere, and the corresponding LCOGT network consists of telescopes at two sites operational during all of 2017: ``TFN'' (Canary Islands) and ``OGG'' (Hawaii), and an additional site which became operational in 2017 November, ``ELP'' (Texas).  At each pointing, we requested an observation sequence comprising a series of 2 to 4 exposures in each of the $B$, \rp, \ip\ filters.  Exposure times were 200, 40 and 30 seconds for the $B$, \rp, \ip\ filters, respectively, leading to a total of $\sim 20$ minutes per observation request including overheads for slew time, filter changes, readout, etc.  We intentional defocused by 1~mm on the focal plane in order to minimize the effects of detector systematics in the photometry, however, we note that the defocus command on the 0.4m network sites is not robustly calibrated, leading to somewhat variable point spread functions.  Each observation sequence was requested at a cadence of every 0.5 to 2 hours, where in reality the number of completed observation requests per night varied depending on the availability of the 0.4m telescope network schedule as well as down time due to weather and technical issues.                 

Images are processed by LCOGT automatically with \texttt{BANZAI}\footnote{https://github.com/LCOGT/banzai} and are then transferred to local machines where we extracted the  photometry for each telescope and filter image stacks using AstroimageJ \citep{Collins2017}.   Due to the variable defocus mentioned above, we allowed the aperture radius of 7 pixels to vary by a multiplication factor of 1.25.  The background annulus inner and outer radius was set to 15 and 25 pixels, respectively.  We selected an ensemble of between 4 and 9 stars in the field of view as reference stars.  We inspected light curves generated for the comparison stars along with the science star to ensure stability of the comparison stars. We also assessed whether poor sky conditions affected data quality and discarded images where the flux measurements for all of the comparison stars deviated by more than 3-sigma from their mean.  We computed nightly averages in each filter for each LCOGT station by taking the weighted mean of the individual measurements.  We then determined the measurement errors by fitting a flat curve to a weeks worth of measurements taken during an inactive period. The combined nightly error was then scaled appropriately assuming a reduced $\chi^2$ of unity.   

We made configuration changes to our observation request on 4 different dates which resulted in small relative flux offsets for the comparison stars and the science target.  These changes occurred on JD 24557932, 24557960, 2457970, 24558044, and marked by gray vertical lines in Fig. \ref{fig:LC}. As discussed in section \ref{sec:wholefit}, the small offsets do not have large impact on our results.

\section{Wavelength Dependence of Light Curve} \label{sec:colors}

We examine how the wavelength dependence changes over the course of the \elsie\ dip family event from May 2017 to December 2017, which includes the $\sim0.5$ \% brightening ``blip'' that occurred afterward the last dip, starting from day $\sim$230 and lasting at least through day 270. However, there is dimming occurring on different timescales that needs to be taken into account: the dips of up to $\sim3$\% lasting 1-2 weeks and small, slow $<1\%$ variation over the entire 7 months of data. As labeled in fig.~\ref{fig:LC}, the dips in order are \elsie, \celeste, \skara, and \angkor. Between the named dips, the flux does not return to pre-\elsie\ levels, indicating a probable secular dimming occurring with the dips. In particular, between \celeste\ and \skara, the flux in the $i'$ band did not return to pre-\elsie\ level but remains steadily at $\sim0.5\%$ below those levels. After the last 2017 dip, \angkor, a small brightening above pre-\elsie\ levels occurs for a few weeks reaching $\sim1\%$ in $B$ band near day 250 (at the point marked ``blip'' in fig.~\ref{fig:LC}). 
However, this long dimming and the blip are only marginal detections (2-3$\sigma$) so this long term trend may not be real.

If there is significant slower secular dimming occurring simultaneously to the dips with a different wavelength dependence, then the measured wavelength dependence of the combined dimming would be a mixture of the wavelength dependences of the dip and the secular dimming. For example, including gray secular dimming will make measurements of the total wavelength dependence of a dip compared to baseline levels appear weaker than it really is, whereas secular dimming with strong wavelength dependence would make the dips' dependences appear stronger. This makes the choice of how to normalize important as it changes the amount of secular dimming included. To minimize these effects, we also examine each dip individually normalized in such a way that any secular dimming is removed.

\subsection{Wavelength dependence over time}\label{sec:wholefit}

\subsubsection{Light Curve Model}
We perform a least squares fit to the entire 2017 light curve with a similar fitting algorithm to the one used to fit the \elsie\ dip in \citet{Boyajian2018} with the only difference being that we allow segments of the light curve to have different wavelength dependence so we can study how the dependence changes with time. Described in detail below, the free parameters we fit together are the depth in the $i'$ band for each day there is data ($d_n$), the depth ratios between the filter, $X$, over a given region ($X/i'$), and the normalization of the entire light curve for each camera in each filter ($N_X$).

We define a depth in the $i'$ band for day $n$ as $d_n=\Delta F_{i'}=N_{i'}-F_{i'}$ and require the depths to be positive. The requirement $d_n>0$ defines $N_{i'}$ to be the highest flux value which occurs near the top of the `blip' around day 255. The depth ratio is then defined as $X/i'\equiv\Delta F_X / \Delta F_{i'}$ where $\Delta F_X = N_X-F_X$ and $N_X$ and $F_X$ are the normalization and flux in either the $B$ or $r'$ band. We include flat normalizations for each telescope camera in each filter as free parameters to take into account the small instrumental differences without removing any long-term trends.
As done with the wavelength dependence fit with \elsie\ in \citet{Boyajian2018}, each day has a depth in the $i'$ filter ($d_n$) that is fit to the data in all the filters with appropriate color and normalization scaling such that $N_X$, $d_n$ and $X/i'$ are fit simultaneously.
The residual we minimize is
\begin{equation}
\chi^2=\sum_{F_X} \frac{\left(F_{X}-N_{X} \times [1-d_n\times X/i']\right)^2}{(\sigma_{F_X})^2}
\label{eq:min}
\end{equation}
where $\sigma_{F_X}$ is the reported uncertainty on the photometry and $X/i'=1$ with $i'$ data. The normalizations, $N_X$, are held constant in time while $d_n$ and $X/i'$ are allowed to vary in time with different binning. The renormalized data are shown with the light curve fit in the top plot of fig. \ref{fig:LC}.

To examine how the wavelength dependence of the dimming changes over time, we define regions of constant $B/i'$ and $r'/i'$ which range from $\sim5-10$ days long. Boundaries are placed at the beginning, middle, and end of each dip and at approximately equal spaces between dips, as well as at the beginning and end of the light curve. These boundaries for separate depth ratio regions are indicated by light gray, vertical dotted lines in Fig \ref{fig:LC} with the \elsie, \celeste, \skara, and \angkor\ dips highlighted in red, yellow, green and blue. The measured $B/i'$ and $r'/i'$  are displayed in the bottom plot of fig. \ref{fig:LC}. The depth ratios for the dips (the colored regions) are listed in table \ref{tab:cols} as total depth ratio, $(X/i')_\mathrm{tot}$, and the depth ratios for the secular dimming (the uncolored regions) in the regions adjacent to dips are listed as secular depth ratio, $(X/i')_\mathrm{sec}$. These regions are labeled 1-17 in Fig.\ref{fig:LC} and their depth ratio values labeled in table \ref{tab:cols}. These labels are explained in section \ref{sec:dips}.


There are small discontinuities in the normalizations as seen by eye in the comparison stars due to instrumental effects. To check the effect of the normalization discontinuities on determining the wavelength dependence, we place additional boundaries in $B/i'$ and $r'/i'$ at the known discontinuities. These boundaries are marked by solid, dark gray vertical lines in fig. \ref{fig:LC}. One of these normalization discontinuities occurs during \skara, so this dip has three color regions instead of two like the others. For all but one, there is no significant change in depth ratio across the discontinuities, demonstrating our choice in normalization is correct. 

\begin{deluxetable*}{lcccc}
\tablecolumns{5}
\tabletypesize{\footnotesize}
\tablecaption{\elsie\ Dip Family Depth Ratios \label{tab:cols}}
\tablehead{\colhead{}	& \colhead{Elsie}	& \colhead{Celeste}	& \colhead{Skara Brae} & \colhead{Angkor}
}

\startdata
\multicolumn{5}{l}{total depth ratio}		 \\
$(B/i')_\mathrm{tot}$		& $2.80\pm0.24$ (2)	& 2.92$\pm0.32$	(6)	& 1.95$\pm0.34$ (10)	& 1.70$\pm0.14$	(15)	\\
			& $3.01\pm0.26$ (3)	& 2.24$\pm0.19$ (7)	& 2.00$\pm0.22$ (11)	& 1.66$\pm0.19$ (16)	\\
            &				&						& 2.05$\pm0.20$ (12)	&			\\
$(r'/i')_\mathrm{tot}$ 	& 1.77$\pm0.15$ (2)	& 1.79$\pm0.20$ (6)	& 1.79$\pm0.30$ (10)	& 1.45$\pm0.12$ (15)	\\
			& 1.83$\pm0.16$ (3)	& 1.84$\pm0.16$ (7)	& 1.63$\pm0.18$ (11)	& 1.60$\pm0.18$ (16) 	\\
			&				&						& 1.65$\pm0.16$ (12)	&			\\
\hline
\multicolumn{5}{l}{dip depth ratio}			  \\
$(B/i')_\mathrm{dip}$	& $1.665_{-0.090}^{+0.090}$	& 3.090$_{-0.171}^{+0.176}$	& 2.309$_{-0.135}^{+0.117}$	& 1.803$_{-0.110}^{+0.112}$		\\
$(r'/i')_\mathrm{dip}$ 	& 1.292$_{-0.070}^{+0.064}$	& 1.554$_{-0.105}^{+0.109}$	& 1.157$_{-0.077}^{+0.077}$	& 1.131$_{-0.080}^{+0.081}$		\\	
\hline
\multicolumn{5}{l}{secular depth ratio}		\\
$(B/i')_\mathrm{sec}$	& 4.92$\pm$2.47 (1)	& 4.76$\pm$1.04 (5)	& 1.92$\pm$0.30 (9)	& 1.29$\pm$0.20 (14)	\\
		& 5.15$\pm$1.23 (4)	& 1.63$\pm$0.27 (8)	& 2.43$\pm$0.97 (13)	&1.27$\pm$0.42 (17)	\\
$(r'/i')_\mathrm{sec}$	& 2.94$\pm$1.46 (1)	& 2.84$\pm$0.61 (5)	& 1.71$\pm$2.59 (9)	& 1.75$\pm$0.23 (14)	\\
		& 2.81$\pm$0.66 (4)	& 1.75$\pm$0.28 (8)	& 3.78$\pm$1.52 (13)	& 2.41$\pm$0.71 (17)	\\
\enddata
\tablecomments{The top values are the total depth ratio fits for each of the two in-dip regions (3 for \skara) that include the slow, secular dimming, $(X/i')_\mathrm{tot}$, as described in section \ref{sec:wholefit} and shown in Fig. \ref{fig:LC}. The middle values are the dip depth ratio fits of only the dips, $(X/i')_\mathrm{dip}$, see section \ref{sec:dips} and Fig. \ref{fig:dips}. The bottom values are the secular depth ratios, $(X/i')_\mathrm{sec}$, see section \ref{sec:wholefit}. The number in the parenthesis refer to regions marked in Fig. \ref{fig:LC}.
}
\end{deluxetable*}

\begin{figure*}
\centering
\includegraphics[width=7.0in, trim= 0 0 0 0 ]{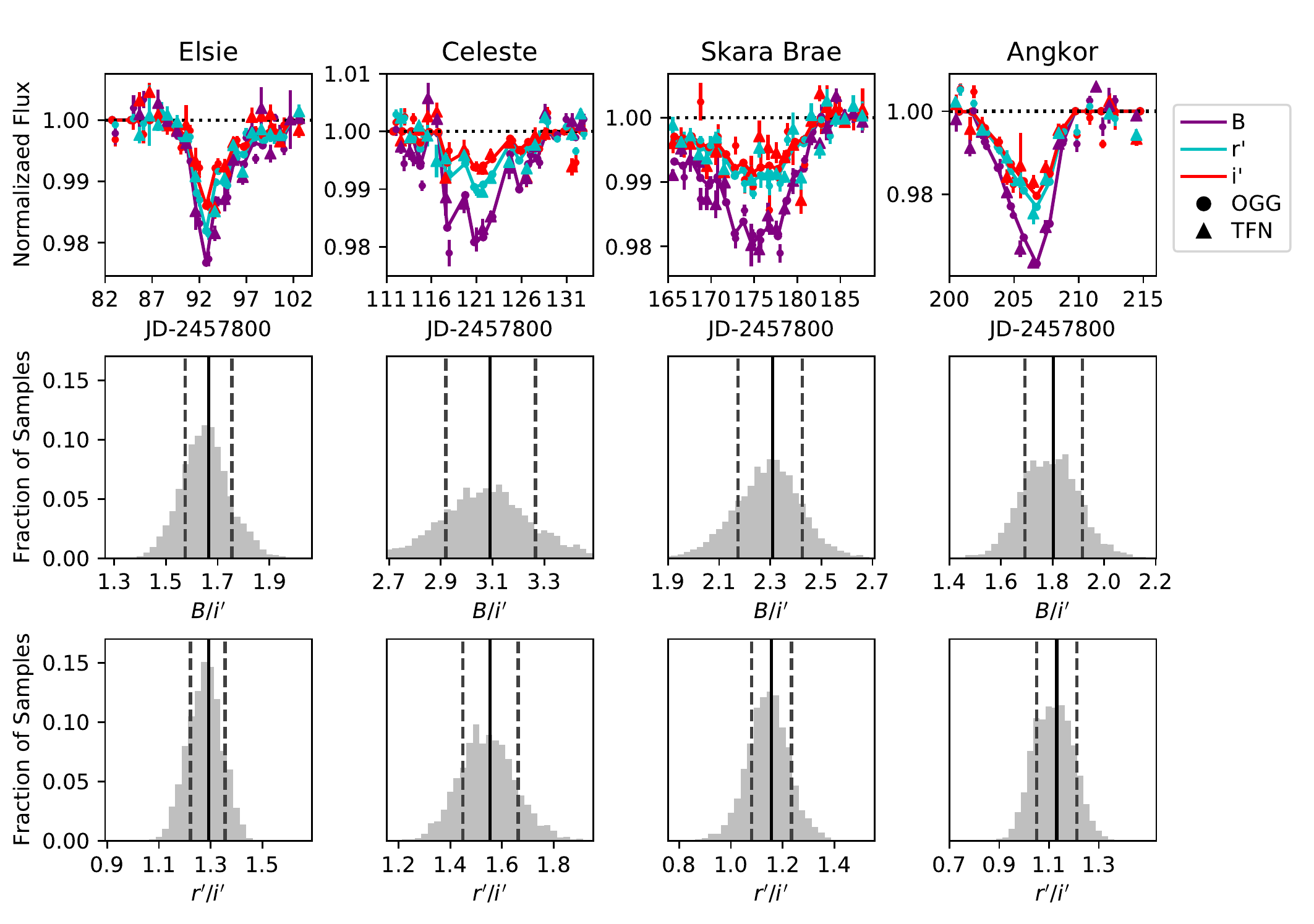}
\caption{The top panels are fits of the normalized dip segments of the light curve normalized with the secular dimming removed as described in section \ref{sec:dips}. These panels are plotted with the same colors and markers as the top plot of fig. \ref{fig:LC}. Below are histograms of the posterior samples for $(B/i')_\mathrm{dip}$ and $(r'/i')_\mathrm{dip}$ from the Bayesian analysis with the average (solid vertical line) and 1$\sigma$ errors (dotted vertical lines).
}
\label{fig:dips}
\end{figure*}

\begin{figure}
\centering
\includegraphics[width=3.45 in, trim= 0 0 0 0 ]{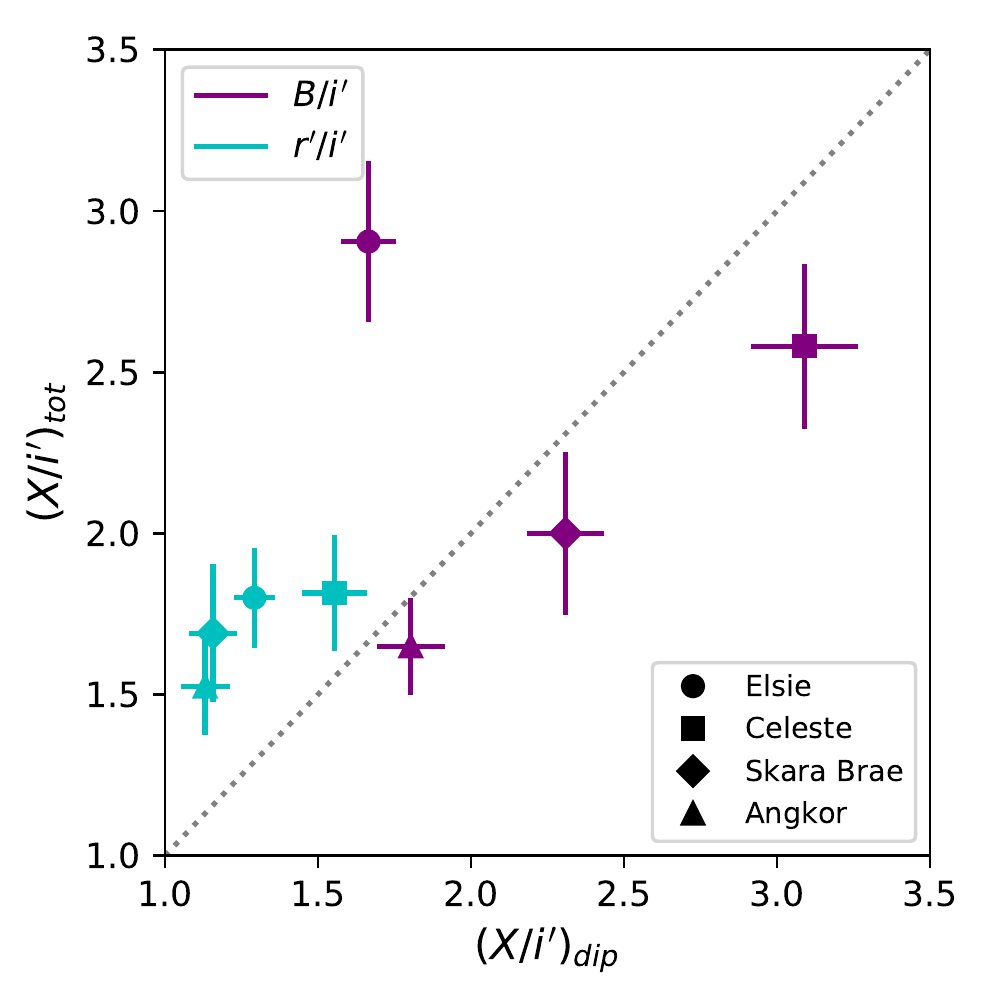}
\caption{Plot comparing of the total depth ratio (dip plus any long term color trends), $(X/i')_\mathrm{tot}$, measured in section \ref{sec:wholefit} versus just the dip depth ratio, $(X/i')_\mathrm{dip}$, as measured in section \ref{sec:dips}. The dotted line marks $(X/i')_\mathrm{tot}=(X/i')_\mathrm{dip}$. For \celeste, \skara, and \angkor, the depth ratios are consistent to within $\sim2\sigma$. The measurements for \elsie\ are inconsistent at the 3$\sigma$ level for $r'/i'$ and 5$\sigma$ for $B/i'$ (see section \ref{sec:dips} for details).}
\label{fig:compare}
\end{figure}

\subsubsection{Results}
The measured depth ratios, $X/i'$, vary from 1 to 6 with the most noticeable changes occurring at some of the boundaries of the dips. The jump in depth ratio from going in to out of dip is particularly large around \elsie\ where $B/i'$ is nearly twice as large in the regions adjacent to the dip than in the dip. However, when $d_n$ is very small, the errors become larger so the differences between in-dip \elsie\ and the regions right before and after it are $<1\sigma$ and 1.5$\sigma$ respectively. Restricting ourselves to regions with small errors ($\lesssim0.2$), $B/i'$ varies from $\sim3 - 1.5$ while $r'/i'$ remains between $\sim2-1.2$.

The variation in depth ratios is less when comparing only the dip regions, (the colored regions in Fig. \ref{fig:LC}). The first two dips, \elsie\ and \celeste\ have $B/i'\approx 3$ but for \skara\ and \angkor\ $B/i'\approx2$. The variation is less for $r'/i'$ with \elsie\ and \celeste\ with $r'/i'\approx1.8$ and \skara\ and \angkor\ with $r'/i'\approx1.6$. 
The largest $B/i'$ region at 3.01$\pm$0.26 occurs during \elsie\ which is inconsistent with the previously reported value 1.94$\pm$0.06 \citep{Boyajian2018} at the $4\sigma$ level. The other \elsie\ $B/i'$ region is also inconsistent with the previous measurement at $B/i'=2.80\pm0.24$. For $r'/i'$, the values measure here, 1.77$\pm$0.15 and 1.83$\pm$0.16 are larger than the previously reported value, 1.31$\pm$0.04 \citep{Boyajian2018} but with only a $3\sigma$ discrepancy. This discrepancy is caused by the strong, secular dimming wavelength dependence contributing to our measured depth ratio by our choice in normalization. When the secular and dip components are separated, the discrepancy reduces. This is discussed further in section \ref{sec:mix}. 

For the secular dimming (the uncolored regions of Fig. \ref{fig:LC}), the $(B/i')_\mathrm{sec}$ and $(r'/i')_\mathrm{sec}$ ratio also changes with time. At the beginning of the light curve, $B/i'\approx$5 and $r'/i'\approx$2.75 and then both decrease to $\sim$1.5. After day 220, the depth ratios become unreliable because of the small depths since the ratios are undefined when $d_n=0$. 

\begin{figure*}
\centering
\includegraphics[width=6.0in, trim= 0 0 0 0 ]{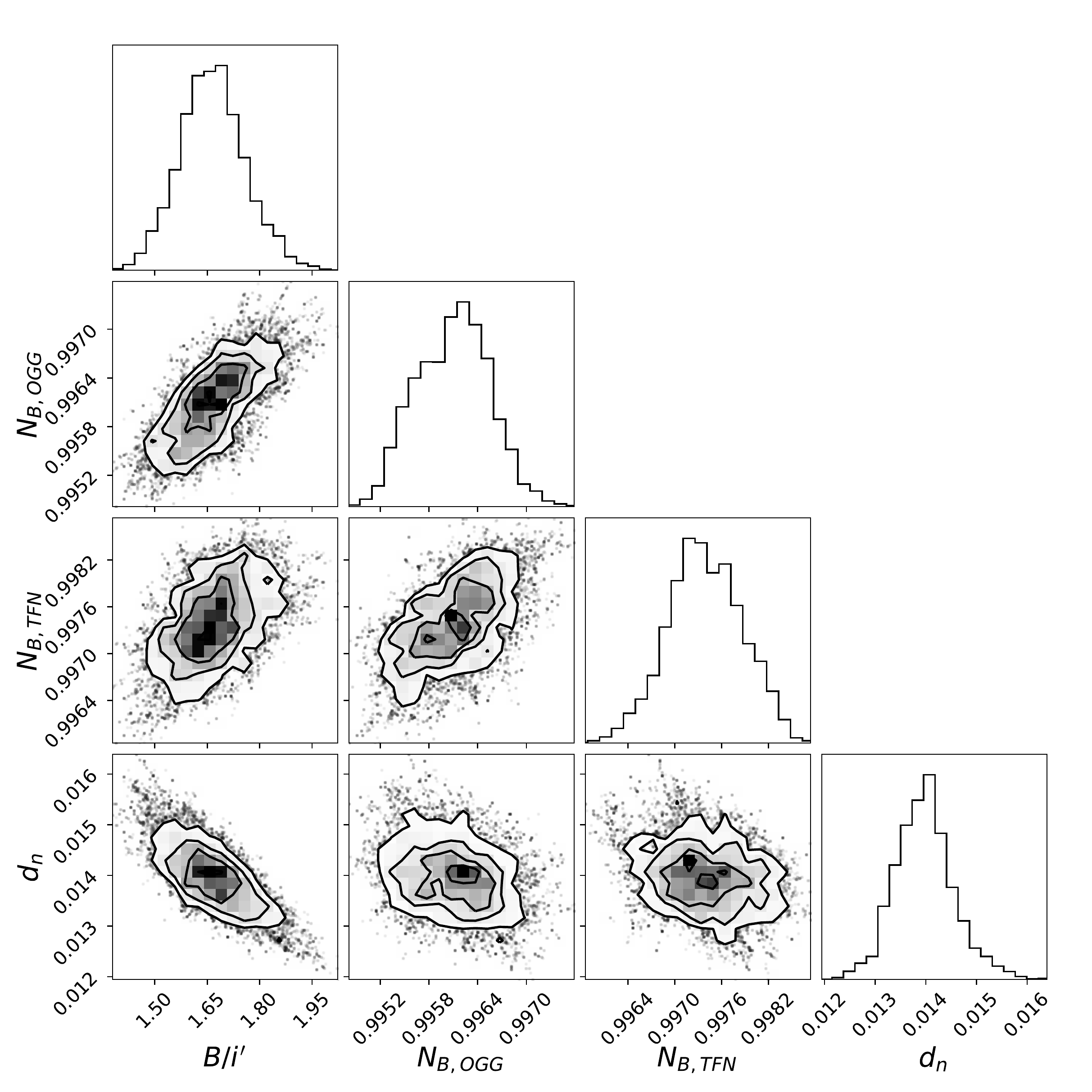}
\caption{Corner plot of the \elsie\ fit with the normalization in $B$ band and $B/i'$ color and $d_n$ for the deepest day of the dip. The 2-D histograms show correlations between $d_n$ and $B/i'$, both $N_B$, and $B/i'$, and between the two $N_B$.
}
\label{fig:Ebcorner}
\end{figure*}

\begin{figure*}
\centering
\includegraphics[width=6.0in, trim= 0 0 0 0 ]{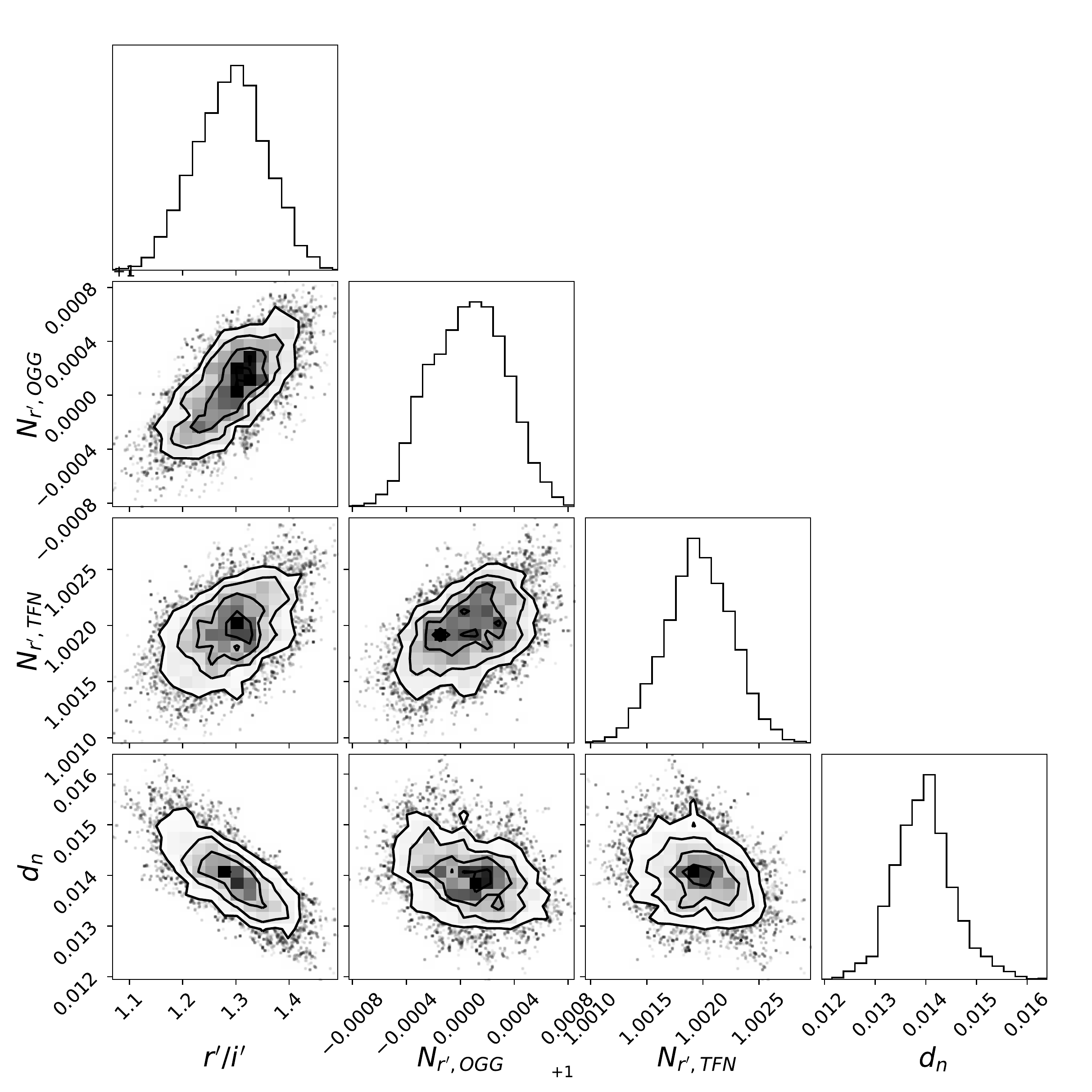}
\caption{Same corner plot of the \elsie\ dip as in fig. \ref{fig:Ebcorner} but for the normalization in $r'$ band and $r'/i'$ color. Same correlated errors occur in the $r'$ band as the $B$ band.
}
\label{fig:Ercorner}
\end{figure*}

\begin{figure}
\centering
\includegraphics[width=3.45 in, trim= 0 0 0 0 ]{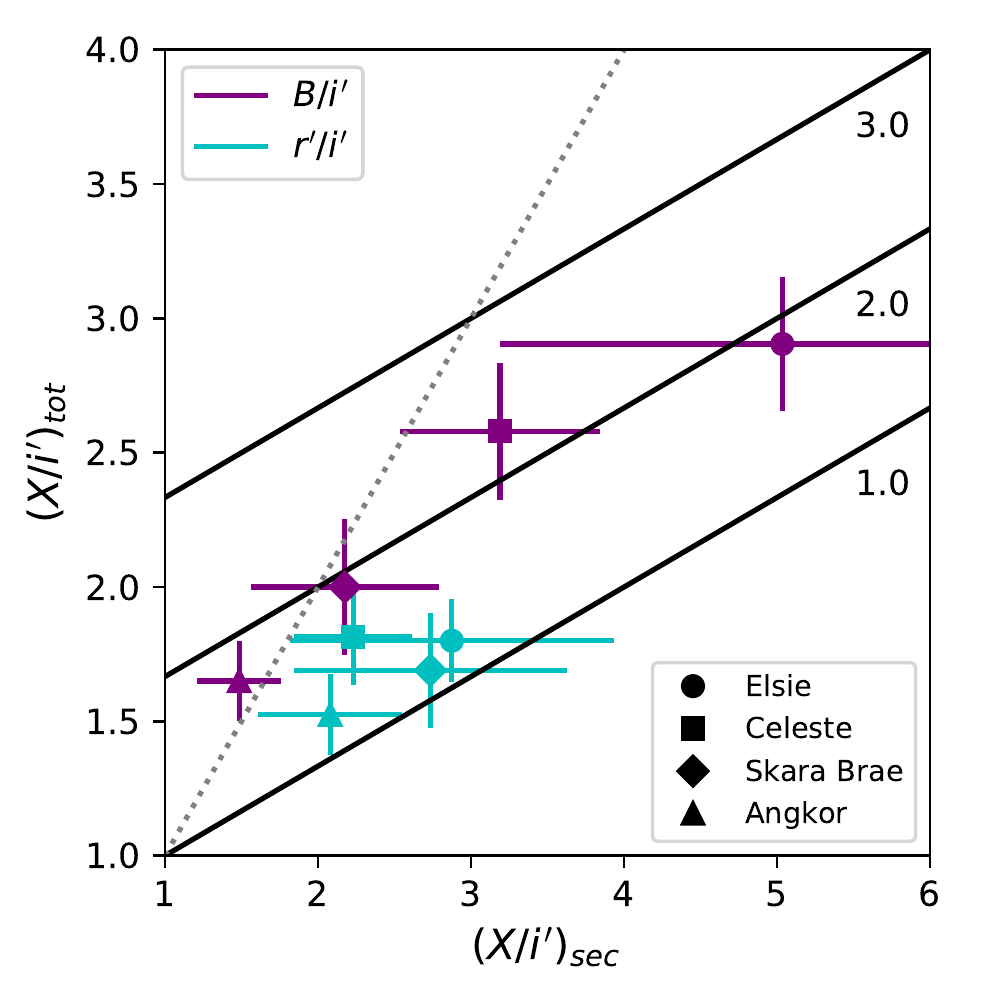}
\caption{Plot of the total depth ratio versus the secular depth ratio with solid black lines of constant dip depth ratio as calculated by equation \ref{eq:deprat} and described in section \ref{sec:mix}. We set the dip to secular extinction ratio in $i'$, $\tau_{i',\,\mathrm{dip}}/\tau_{i',\,\mathrm{sec}} =2$. The dotted line marks $(X/i')_\mathrm{tot}=(X/i')_\mathrm{sec}=(X/i')_\mathrm{dip}$.  If the dip-to-background-depth ratio increases (decreases), then the slope decreases (increases). For values of $(X/i')_\mathrm{sec}$ on the right side of the dotted line, the measured depth ratio with will be stronger than if the secular dimming is removed. The average observed depth ratio for each dip listed in table \ref{tab:cols} with the average of the immediately adjacent background depth ratio regions are plotted in purple for $B/i'$ and cyan for $r'/i'$. The dips are marked by different shapes. Dip depth ratios of $(X/i')_\mathrm{dip}\sim2$ and $\sim1.5$ are favored for $B/i'$ and $r'/i'$, respectively which confirms our assumption the secular dimming and the dips are caused by multiple optically-thin dust populations that overlap along the line of sight and can be treated separately. 
}
\label{fig:mix}
\end{figure}

\begin{figure*}
\centering
\includegraphics[width=7.0in, trim= 0 0 0 0 ]{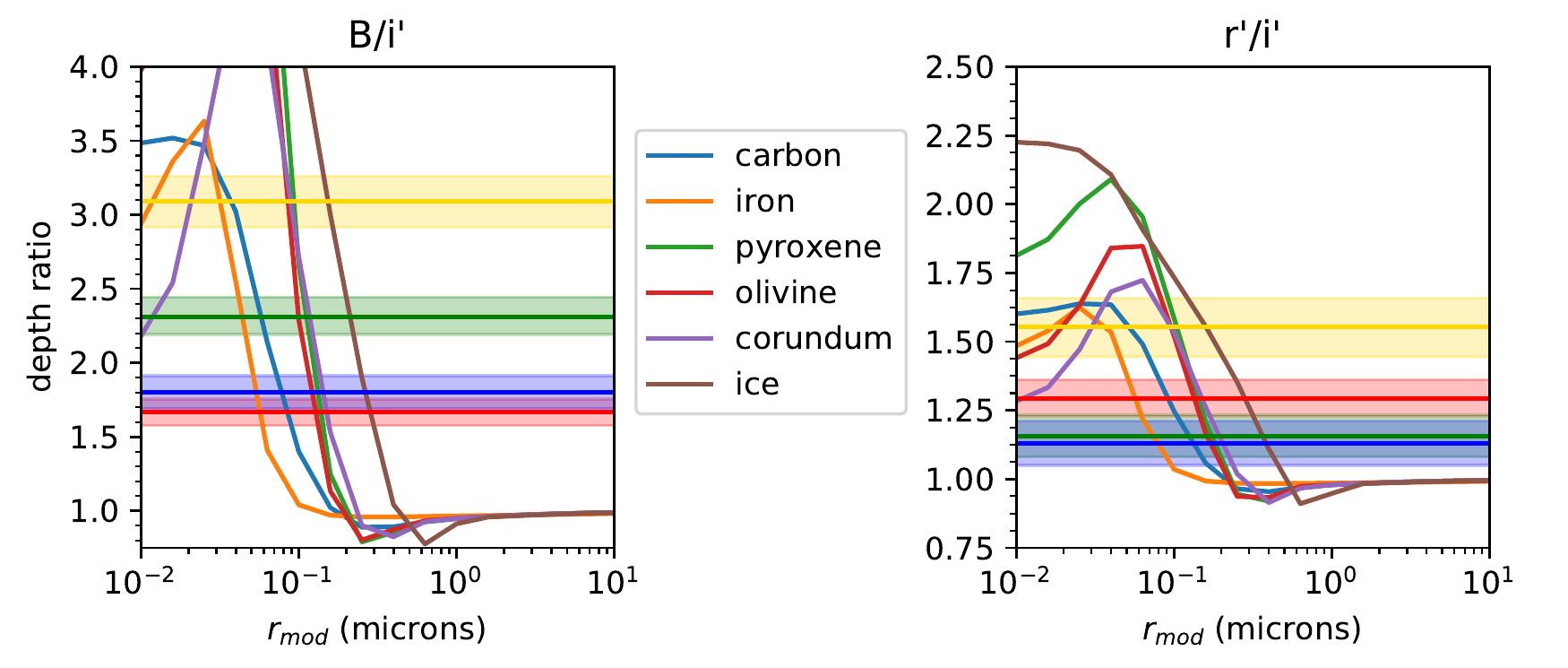}
\caption{Depth ratios for a given average grain size estimated for six different dust grain compositions. The observed dip depth ratios, $(X/i')_\mathrm{tot}$, for each dip as described in section \ref{sec:dips} are shown by red, yellow, green, and blue horizontal lines for \elsie, \celeste, \skara, and \angkor\ (the same color scheme as in figure \ref{fig:LC}). Despite the large difference in depth ratios, all of the dips have a similar dust grain size range, $<0.5\,\mu m$.
}
\label{fig:ratios}
\end{figure*}

\subsection{Color of the Dips}\label{sec:dips}

To remove the effects of the secular wavelength dependence, we look at segments of the of the light curve just a few days wider than each individual dip. We denote the depth ratios measured for the colored regions in section \ref{sec:wholefit} as the ``total depth ratios'' $(X/i')_\mathrm{tot}$ because we did not remove any long-term color trends from the dip depths. The depth ratios measured in this section are denoted ``dip depth ratios'' $(X/i')_\mathrm{dip}$ since the secular dimming is removed. The relationship between these two ratios, $(X/i')_\mathrm{tot}$ and $(X/i')_\mathrm{dip}$ is discussed in more detail in section \ref{sec:mix}. 

We perform the fitting method to the dip light curve segments but artificially set the normalization of the light curve to the level just prior to or just after the dip (or both in the case of \celeste) depending whichever level is higher, by setting $d_n=0$ for $5-6$ days in the pre- or post-dip region.
The least squares fit that resulted in an adequate solution, but the normalizations had a tendency to increase with decreasing wavelength, which reduces the flux values in $B$ and $r'$ bands such that the flux is no longer normalized to one just outside the dip. The lower fluxes artificially increase the depth ratios. To overcome that degeneracy in the fit, we explored the parameter space with the Markov chain Monte Carlo method with the \textit{emcee} code \citep{Foreman2013}. We use flat priors over ranges 0.5-5, 0.99-1.01, and $>$0 for the depth ratios, normalizations, and $d_n$, respectively, and set equation \ref{eq:min} as the likelihood with the results of the least squares fit as the initialization for the MCMC sampler.
The results of the Bayesian fit and the posteriors of the depth ratios, $(X/i')_\mathrm{dip}$ are shown in fig. \ref{fig:dips} with the averages and $1\sigma$ confidence levels listed in Table \ref{tab:cols}.

After removing the background wavelength dependence, $(X/i')_\mathrm{dip}$ except for those of \elsie\ remain fairly consistent with $(X/i')_\mathrm{tot}$ from the whole light curve fit as shown in fig. \ref{fig:compare}.
 For \celeste, \skara, and \angkor\, $B/i'$ are each consistent within $\lesssim\sigma$ but $r'/i'$ is lower for each dip by $\lesssim2\sigma$.
However, the depth ratios for \elsie, $(B/i')_\mathrm{dip}=1.67$ and $(r'/i')_\mathrm{dip}=1.29$ are significantly smaller than the total depth ratio values, $(B/i')_\mathrm{tot}\approx3$ and $(r'/i')_\mathrm{tot}\approx1.8$ with 5 and 3$\sigma$ discrepancies, respectively. The effects of including the secular dimming are stronger when the difference in wavelength dependence is larger and \elsie\ has the largest jump in depth ratio going from in- to out-of-dip. The \elsie\ depth ratios from this renormalized fit are slightly smaller than the previously reported value from \citet{Boyajian2018} but the discrepancy is only 2$\sigma$ for $B/i'$ and $<1\sigma$ for $r'/i'$. This small discrepancy is due to a different choice in the number of out-of-dip days were used to normalize the light curve.  For \celeste, the second $(B/i')_\mathrm{tot}$ region (days $\sim120-130$) is inconsistent with the $(B/i')_\mathrm{dip}$ value and $\sim4\sigma$ lower. This low $B/i'$ region is a shallower part of the dip where wavelength dependence is dominated by the lower secular depth ratio value. 

The fits have a small tendency to favor larger depth ratios, which is reflected in the slight asymmetry in the depth ratio posteriors in fig. \ref{fig:dips}. The normalization of a filter is also fairly correlated with depth ratio for that filter as shown in the top two 2-D histograms in the corner plots for \elsie\ displayed in figs. \ref{fig:Ebcorner} and \ref{fig:Ercorner}. This causes some of the normalizations to have the same asymmetry. Despite $d_n$ being compared to data in all filters, $d_n$ is strongly correlated to $X/i'$ in the same direction for both $B$ and $r'$ band. This pattern is true for the other dips except for $B$ band with \celeste. \celeste\ is much shallower than the rest after renormalization so the errors are larger.

\subsection{Blending Wavelength Dependences}\label{sec:mix}

Assuming the secular dimming and the dips are from two separate, optically-thin dust populations, then the total extinction $\tau_\mathrm{tot}$ is the sum of the opacities of each dust population, i.e., $\tau_\mathrm{tot}=\tau_\mathrm{sec}+\tau_\mathrm{dip}$. 
Since the extinction is approximately the dip depth in the optically thin regime ($d_n\approx\tau_{i'}$), the depth ratio during a dip with background dimming is
\begin{eqnarray}
\left(\frac{X}{i'}\right)_\mathrm{tot} &=& \frac{\tau_{X,\,\mathrm{sec}}+\tau_{X,\,\mathrm{dip}}}{\tau_{i',\,\mathrm{sec}}+\tau_{i',\,\mathrm{dip}}}  \nonumber\\
&=& \frac{(X/i')_\mathrm{sec}}{1+\frac{\tau_{i',\,\mathrm{dip}}}{\tau_{i',\,\mathrm{sec}}}}+\frac{(X/i')_\mathrm{dip}}{1+\frac{\tau_{i', \,\mathrm{sec}}}{\tau_{i',\, \mathrm{dip}}}}
\label{eq:deprat}
\end{eqnarray}
where $X$ is the $B$ or $r'$ filter. The measured depth ratio depends on three factors: the depth ratio of the secular dimming material, the depth ratio of the dip material and the ratio of extinction from the dip to the secular material. For the secular material to have an impact on the measured depth ratio ($(X/i')_\mathrm{tot}$), two criteria must be met: the depth ratios of the secular material and and the dip material must be significantly different ($\vert (X/i')_\mathrm{sec}-(X/i')_\mathrm{dip}\vert \gg 0 $) and the extinction due to secular dimming must be comparable to the dip extinction ($\tau_{i',\,\mathrm{dip}}/\tau_{i',\,\mathrm{sec}} \sim1$).

In figure \ref{fig:mix}, we show lines of constant $(X/i')_\mathrm{dip}$ assuming $\tau_{i'\mathrm{dip.}}/\tau_{i'\mathrm{sec}} = 2.0$ on a plot of $(X/i')_\mathrm{tot}$ versus $(X/i')_\mathrm{sec}$. The line of equal depth ratios, $(X/i')_\mathrm{tot}=(X/i')_\mathrm{sec}=(X/i')_\mathrm{dip}$ is marked by the dotted line. 
The slopes of constant $(X/i')_\mathrm{dip}$ decrease (increase) with increasing (decreasing) $\tau_{i'\mathrm{dip}}/\tau_{i'\mathrm{sec}}$ and pivot at the point of equal depth ratios.
On the right side of the plot, the secular dimming has a stronger influence on the wavelength dependence and the dips will appear to have a larger depth ratio than they actually do.

We overlaid the average values for $(X/i')_\mathrm{tot}$ and $(X/i')_\mathrm{sec}$ found in section \ref{sec:wholefit} (see also table \ref{tab:cols}) on fig. \ref{fig:mix}. For the average $(X/i')_\mathrm{sec}$, we used the immediately adjacent regions on each side of the dip. Since the data points are within a couple $\sigma$ of the line of equal depth ratios, the effect of changing the extinction ratio is small compared to the errors. Even though \elsie\ has a much larger observed $(B/i')_\mathrm{tot}$ than Angkor, both are consistent with $(B/i')_\mathrm{dip}=2$ which brings $(B/i')_\mathrm{tot}$ of \elsie\ in agreement with previous measurements. If we changed the reference for normalizing the light curve, $(X/i')_\mathrm{tot}$ and $(X/i')_\mathrm{sec}$ would change but $(X/i')_\mathrm{dip}$ should remain the same.

At $(B/i')_\mathrm{dip}=3.09$ and $r'/i'\approx1.55$, \celeste\ is much redder than \elsie, \skara\ and \angkor\ where $(B/i')_\mathrm{dip}= 1.67$, 2.31 and 1.80 and $(r'/i')_\mathrm{dip}= 1.29$, 1.16a and 1.13, respectively. In fig. \ref{fig:mix}, the $r'/i'$ points for all the dips fall in the expected 1-1.5 range for dip depth ratio and $B/i'$ points for \elsie, \skara\ and \angkor\ fall in the expected 1.5-2.0 range as well. However, the $(B/i')_\mathrm{dip}$ value for \celeste\  is much higher than expected than what would be expected from this plot. This is because one $(B/i')_\mathrm{tot}$ region is much lower than the other which brings down the average. If only the higher $B/i'$ region is used, then the point would be near the 3.0 dip ratio line and consistent with measured values. This confirms the assumption that the dips and the secular dimming are caused by two optically thin dust populations which can be treated separately is good.

\section{Grain Properties}\label{sec:grains}
Assuming that the dips are caused by transiting clouds of optically-thin dust, we can loosely constrain the grain size and composition of the dust from the wavelength dependence of the dips. We compare our measured depth ratios with the secular dimming effects removed $(X/i')_\mathrm{dip}$ from section \ref{sec:dips} to theoretically calculated dust opacities for optically-thin distributions in fig. \ref{fig:ratios}. The opacities were taken from tables from \citet{Budaj2015} which assume homogeneous spherical dust grains with a Diermendjian size distribution \citep{Deirmen1964}.  Fig. \ref{fig:ratios} is similar to figure 8 in \citet{Boyajian2018} except that we explore fewer compositions, only five refractory dust species and water ice. For silicate compositions which can contain varying amounts iron, we consider one pyroxene mineral (Mg$_{1-x}$Fe$_x$Sio$_3$) with a 60/40 Fe/Mg ratio and one olivine mineral (Mg$_{2-y}$Fe$_{2-2y}$Sio$_4$) with a 50/50 Fe/Mg ratio. Optical properties of these minerals are sensitive to iron content but this has little impact on our analysis of the depth ratios here since there is a composition and grain size degeneracy. We also consider other refractory compositions: carbon, iron, and corundum.

For pyroxene, olivine, and corundum, we find these minerals fit $B/i'$ and $r'/i'$ of all the dips with an average grain size of 0.1-0.3 $\mu$m. Carbon and iron fit with slightly smaller grain sizes, $\lesssim 0.1\,\mu$m, and are $\sim1/2$ as large for \celeste\ than \elsie\ and \angkor. Water ice requires a larger grain size of 0.2-0.4 $\mu$m with the grains for \celeste\ on the small side of the range and \elsie\ on the large. 

From day 81 to 244, the secular dimming wavelength dependence ranges over $B/i'\approx 5.5-1$ and $r'/i' \approx 3-1.2$. The $r/i'$ range goes higher than the theoretical depth ratio curves. However, the regions that are above the curves have large enough errors to be consistent with at least the ice and pyroxene curves. The highest secular $B/i'$ is only $\sim1.5\sigma$, inconsistent with the iron and carbon curves. The theoretical curves for olivine, pyroxene, and ice all reach over $B/i'=5.5$. With the large uncertainties of the secular depth ratios, no dust composition can be excluded. The lowest value for $r'/i'$ constrains the grain size to $<0.5\,\mu$m, with ice requiring the largest grains. However, $>1\,\mu$m grains are not excluded at the $2\sigma$ level. $B/i'$ does not provide an upper limit as the lowest value is consistent with gray extinction from large grains.

\section{Discussion}
The variable wavelength dependence observed in the light curve suggests that the dusty material occulting \KIC\ is heterogeneous in size, composition, or both, but small changes in these properties can result in large changes in wavelength dependence. The opacity of a dust grain is very sensitive to the grain size in the sub-micron range for visible light. Despite $(B/i')_\mathrm{dip}$ being much larger for \celeste\, the grain size is only a factor of $\lesssim2$ smaller than the other three dips. The depth ratios can be even more sensitive to the composition. For example, a 0.2 $\mu$m-size grain is consistent with \celeste\ if ice, consistent with \elsie\ and \angkor\ if silicate, or gray if iron. The dust is highly unlikely to be homogeneous in composition, so small changes in relative proportions could explain the variation in wavelength dependence as well as size fluctuations.

For circumstellar dust, grains will be blown out of the system when the radiation pressure is comparable to the gravitational force, e.g. $\beta=0.5$ \citep{Burns1979}. In the \KIC\ system, this occurs for grain sizes of $r\lesssim1\,\mu$m \citep[see Figure 2 in ][]{Neslusan2017}.  For the wavelength dependence measured for each of the dips, we constrain the dust grain size to $r<0.5\,\mu$m so the dust would need to be newly created. The variations in wavelength dependence and small grain size can be naturally explained if dust is brought in with larger objects on elliptical orbits. Both exocomets \citep{Bodman2016} and dust enshrouded planetesimals \citep{Neslusan2017} would form clouds of newly-produced, small-grain dust and variations in composition of the parent bodies would produce a heterogeneous dust cloud with varying properties.
 
The slow secular dimming appears to have a non-gray color dependence for much of the \elsie\ dip family which is in contrast to the gray dimming ($R_V>5$) measured by \citet{Meng2017}. However, due to the larger uncertainties on the measurements of the depth ratios for the secular dimming regions, nearly-gray dimming from large grains is only excluded at the $\lesssim3\sigma$ level.
The uncertainty of the depth ratios also do not include the uncertainty from the choice in normalization and the choice in normalization likely accounts for some of the discrepancy.
If the chosen normalization removes most of the gray secular dimming, then the wavelength dependency will be artificially large.

However if the long-term wavelength dependency is real, then models for the secular dimming that invoke the large-grain, circumstellar dust such as the one proposed by \citet{Wyatt2018} cannot explain all of the long-term dimming occurring with the  \elsie\ dip family. \citet{Wyatt2018} proposed a ring of dust to model the observed century-long secular dimming and the dimming observed during the \textit{Kepler} mission to test the IR excess constraints. This model assumed old dust bound to the system with a grain size $>2\,\mu$m which is much larger than our grain size estimates of $r\lesssim0.5$. The grains we observe causing the secular dimming are too small to remain in the system for many orbits and must be newly created. While their model ignores small grain dust, it is not in conflict some new dust production and not all of the dust producing the secular dimming may be sub-micron. The wavelength dependence of the secular dimming appears to decrease over the extent of the \elsie\ dip family and eventually becomes consistent with gray dimming caused by large grains. As such, the secular dimming may be only partially caused by new, small-grain dust. 



\section{Conclusions}
The 2017 light curve of KIC 8462852 consisting of the \elsie\ dip family has a variable wavelength dependence that changes from dip to dip, suggesting variation in the properties of the dust. There also appears to be a slowly varying wavelength dependence from the secular dimming over the several months in which that dip family occurs. The secular dimming begins with a strong wavelength dependence near \elsie, becoming more gray with time, but the small dip depths cause the depth ratio measurement to have large uncertainties making the non-gray dimming detection only marginal. 

Since the dips and secular dimming occur together, the reference point for normalizing the light curve is important when measuring the wavelength dependence of the dips. The inclusion of long-term dimming with a different wavelength dependence changes the dip depth ratios, but because the secular dimming is smaller, most of the in-dip measurements with and without secular dimming are consistent to $2\sigma$. However, small differences in wavelength dependence can produce large uncertainties in dust grain properties. The largest discrepancy between the two methods is for \elsie, where the strongest secular dimming wavelength dependence occurs. The depth ratio including secular dimming for \elsie is $B/i'\approx3$ and $r'/i'\approx1.8$ but after removing the secular dimming, the depth ratios reduce to$B/i'=$ 1.665 and $r'/i'=1.292$, which are more consistent with previously reported measurements from \citet{Boyajian2018}.

Although the wavelength dependence is variable, almost all the depth ratio measurements are all consistent with sub-micron dust, which agrees with \citet{Deeg2018}. The only exception is the post-\angkor\ secular dimming which is consistent with gray dimming (and hence larger grain size) but the uncertainties are large. For the compositions studied here, the grain size is $r<0.5\,\mu$m, with the largest grains being of water ice and the smallest being carbon with $r<0.1\,\mu$m.  Some of the secular dimming depth ratios are higher ($B/i'>4$ and $r/i'>2$) than calculated for iron and carbon dust models but these compositions cannot be excluded due to large uncertainties in the measurements from the fit measurement and from the uncertainty in the amount of secular dimming. Assuming circumstellar material is causing the dips, both populations of the dust measured in this paper must be newly created, as grains size $r\lesssim1\,\mu$m will be blown out of the system by an early F star.
Ongoing multiband photometric monitoring is needed to better constrain the properties of the dust and the mechanism producing the dust.

Our wavelength dependence measurements do not exclude extinction from the ISM nor other intervening dusty material that is not in the system. More observations such as searches for infrared excess and periodicity are needed to confirm that the dusty material is indeed circumstellar. Intrinsic stellar variations from cooling $\sim$30 K are consistent with the weaker wavelength dependence of \elsie\ and \angkor\ \citep{Foukal2017b}. More work is needed to test if the wavelength dependence of the rest of the light curve is consistent.

\acknowledgments
EHLB's research was supported by an appointment to the NASA Postdoctoral Program within the Nexus for Exoplanet System Science, administered by Universities Space Research Association under contract with NASA. The Center for Exoplanets and Habitable Worlds is supported by the Pennsylvania State University, the Eberly College of Science, and the Pennsylvania Space Grant Consortium. This work makes use of observations from the LCOGT network. The LCOGT observations used in this project were made possible by the collective effort of 1,762 supporters as part of the Kickstarter campaign “The Most Mysterious Star in the
Galaxy” \footnote{\href{https://www.kickstarter.com/projects/608159144/
the-most-mysterious-star-in-the-galaxy/}{\texttt{https://www.kickstarter.com/projects/608159144/
the-most-mysterious-star-in-the-galaxy/}}}. The authors gratefully acknowledge and humbly extend a special thanks for substantial support from Las Cumbres Observatory, Glenn Klakring, Fred Boyajian \& Bobbie Staley, Alex Mazingue, The Bible Family, Claudio Bottaccini, Joachim De Lombaert, Amity \& BrigidWilliams, Kevin Fischer, William Hopkins, Milton Bosch, Zipeng Wang, TJ, DR, and CC.

\bibliography{KICReferences.bib}

\appendix
\section{Corner Plots}\label{sec:extracorner}
We show the corner plots for the fits of \celeste, \skara\ and \angkor\ as described in section \ref{sec:dips}. As seen in fig. \ref{fig:Ebcorner} and \ref{fig:Ercorner}, $d_n$ and $X/i'$ are correlated and $N_X$ and $X/i'$ are correlated.

\begin{figure*}
\centering
\includegraphics[width=7.0in, trim= 0 0 0 0 ]{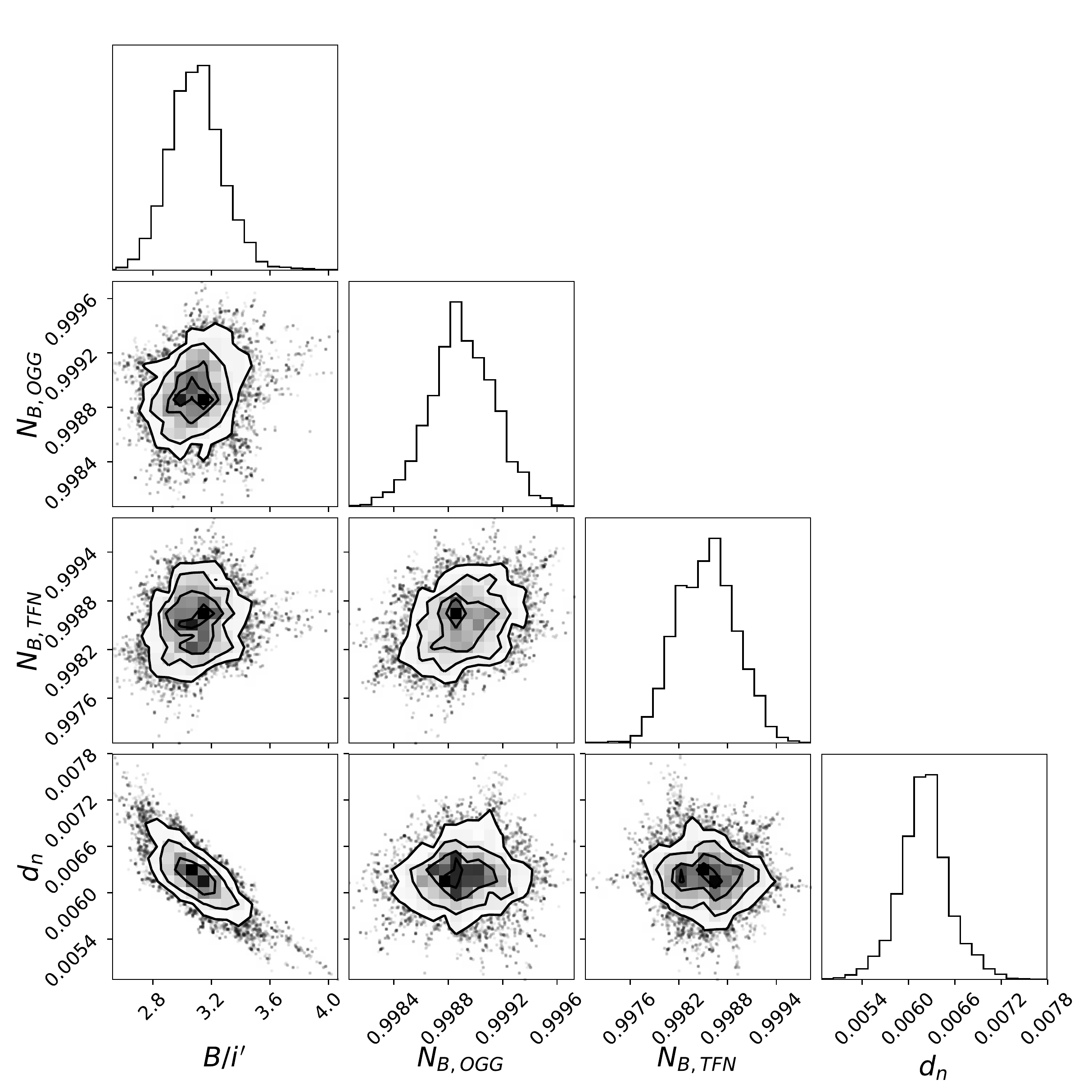}
\caption{Corner plot for the \celeste\ dip for the normalization in $B$ band and $B/i'$ color and the depth for the deepest day.
}
\label{fig:Cbcorner}
\end{figure*}

\begin{figure*}
\centering
\includegraphics[width=7.0in, trim= 0 0 0 0 ]{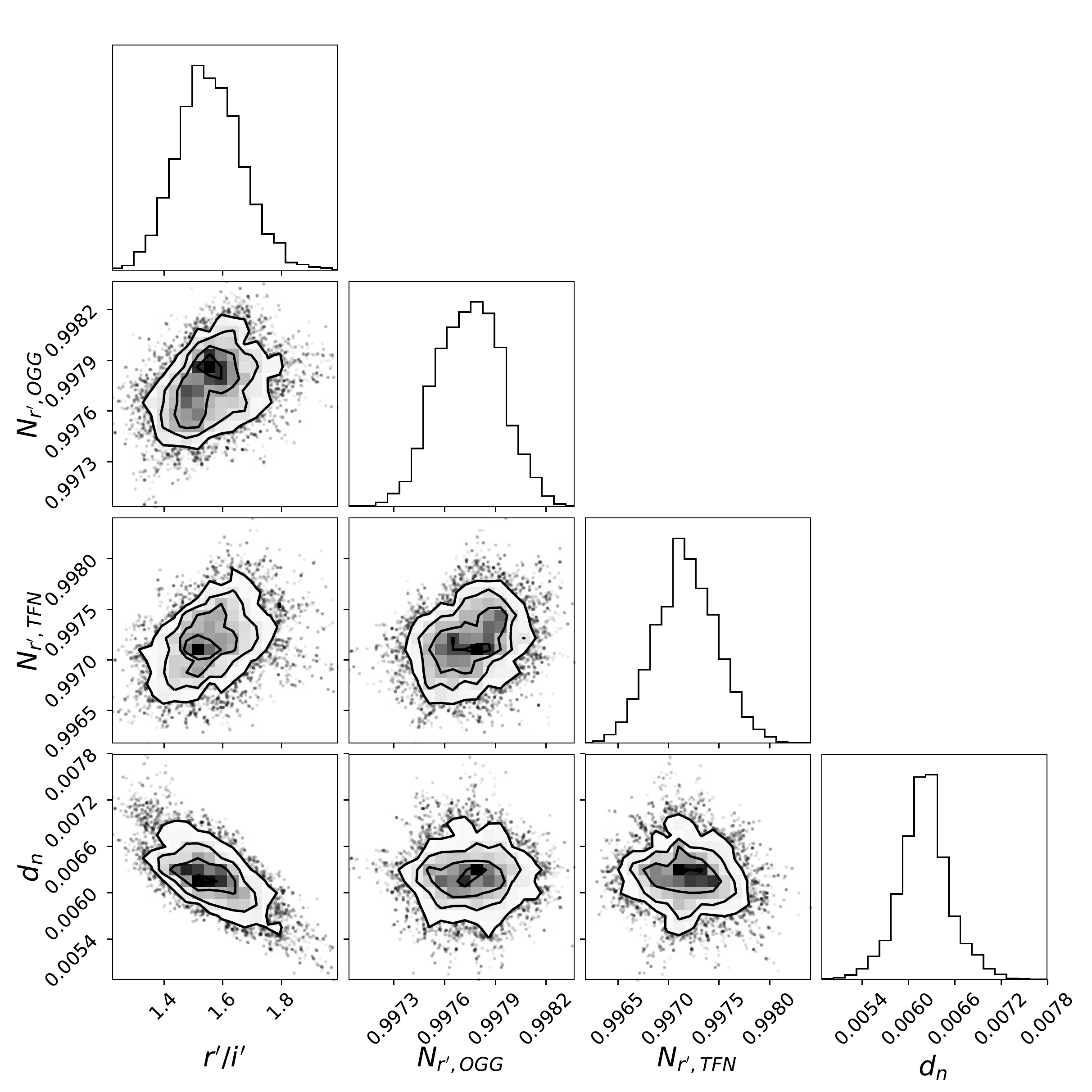}
\caption{Corner plot for the \celeste\ dip for the normalization in $r'$ band and $r'/i'$ color and the depth for the deepest day.
}
\label{fig:Crcorner}
\end{figure*}

\begin{figure*}
\centering
\includegraphics[width=7.0in, trim= 0 0 0 0 ]{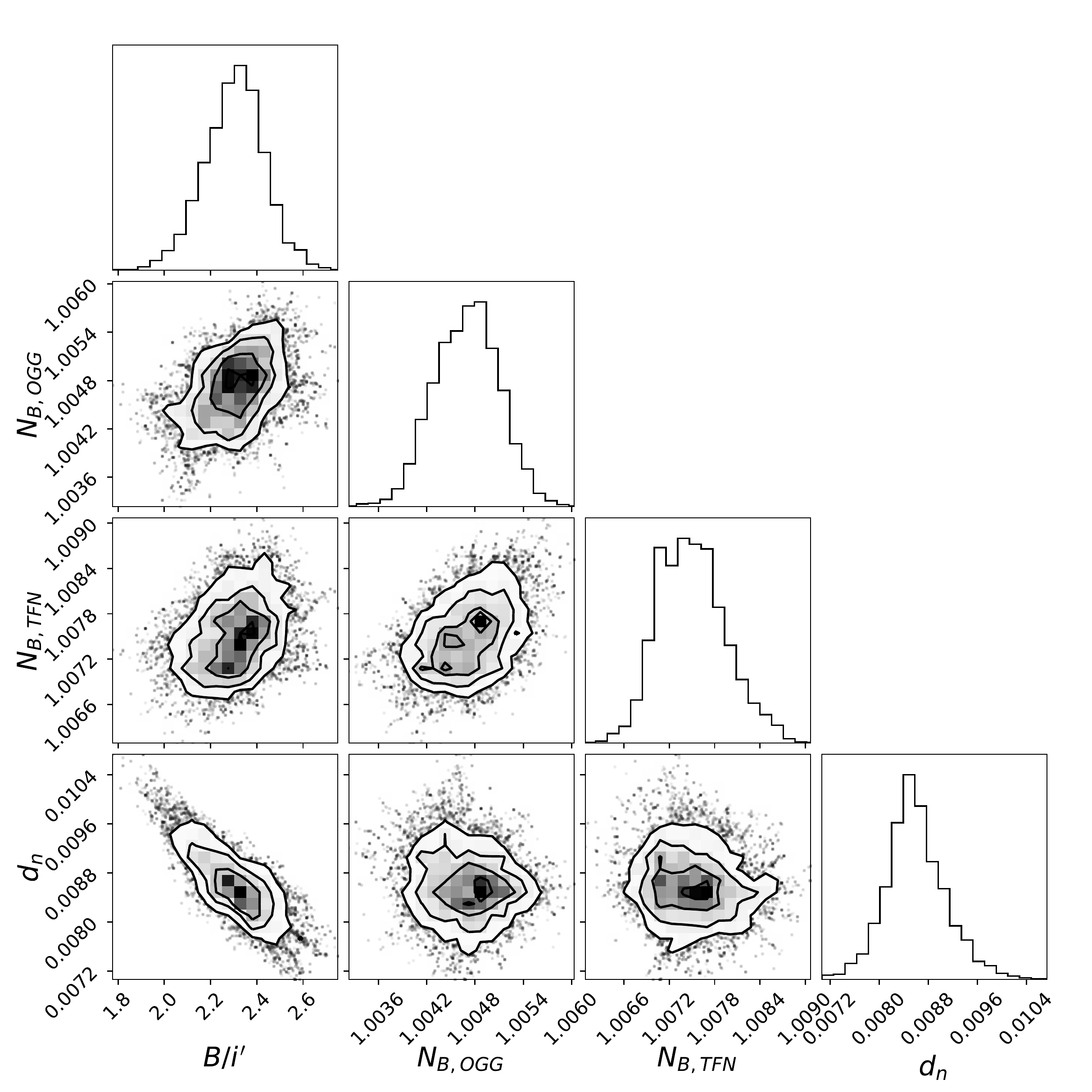}
\caption{Corner plot for the \skara\ dip for the normalization in $B$ band and $B/i'$ color and the depth for the deepest day.
}
\label{fig:Sbcorner}
\end{figure*}

\begin{figure*}
\centering
\includegraphics[width=7.0in, trim= 0 0 0 0 ]{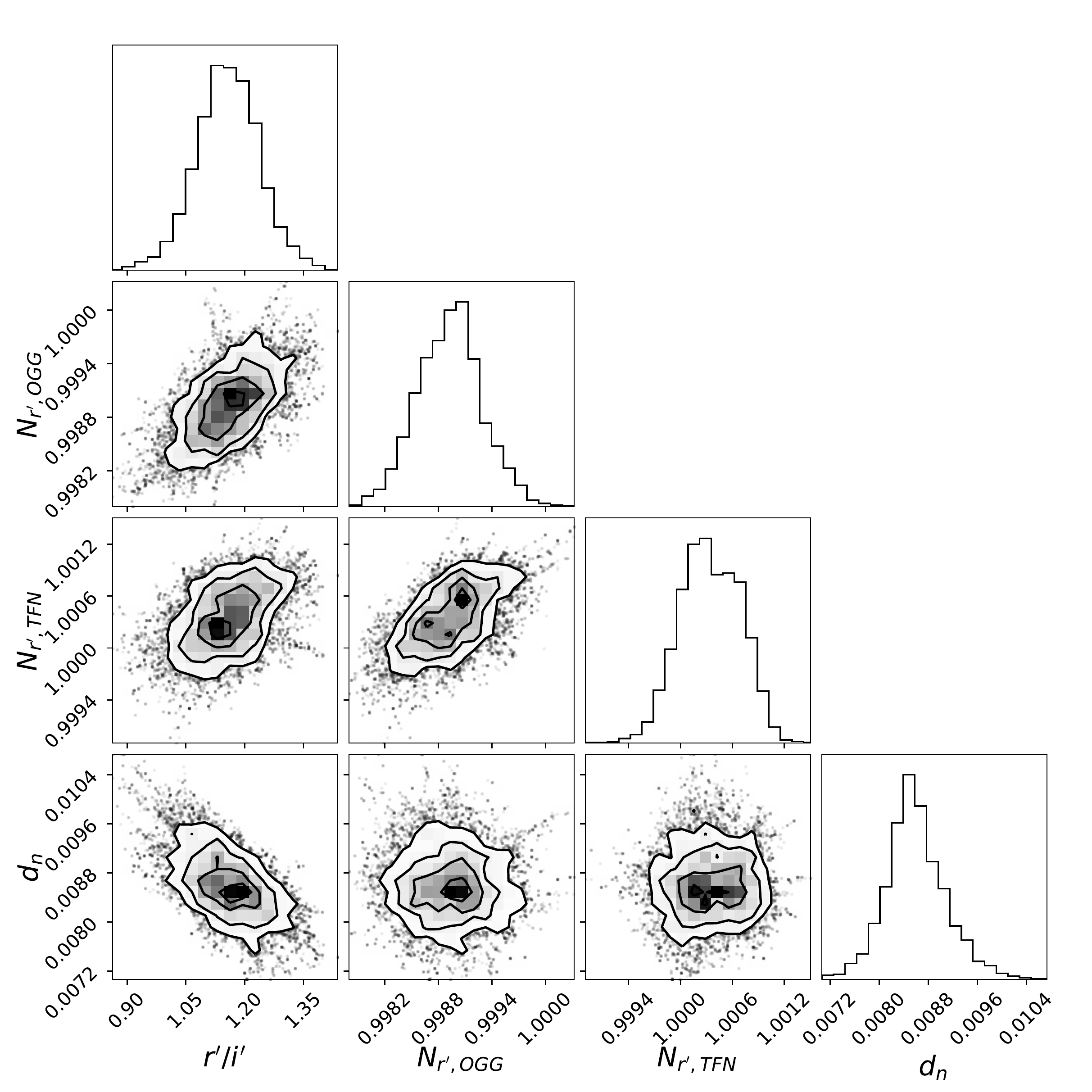}
\caption{Corner plot for the \skara\ dip for the normalization in $r'$ band and $r'/i'$ color and the depth for the deepest day.
}
\label{fig:Srcorner}
\end{figure*}

\begin{figure*}
\centering
\includegraphics[width=7.0in, trim= 0 0 0 0 ]{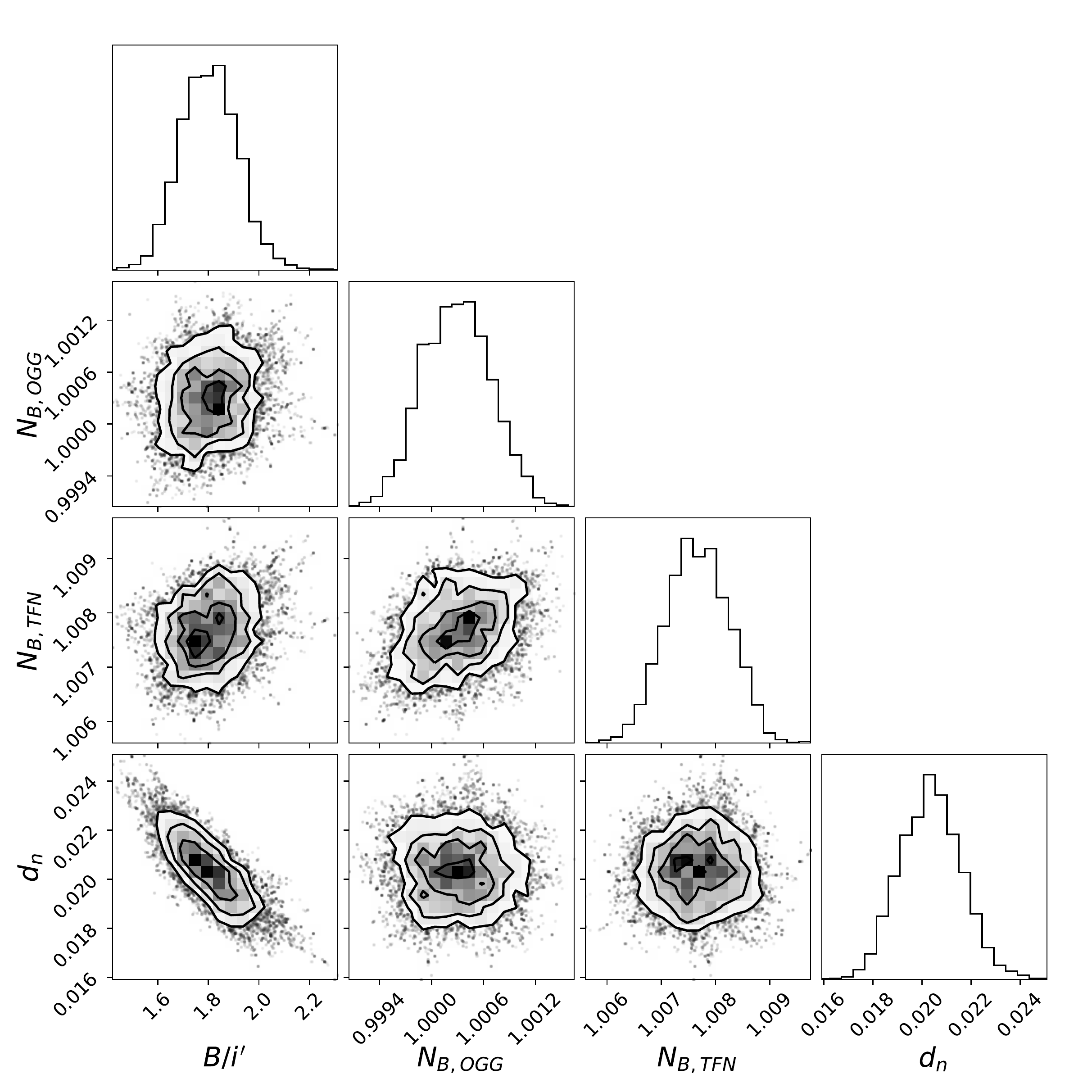}
\caption{Corner plot for the \angkor\ dip for the normalization in $B$ band and $B/i'$ color and the depth for the deepest day.
}
\label{fig:Abcorner}
\end{figure*}

\begin{figure*}
\centering
\includegraphics[width=7.0in, trim= 0 0 0 0 ]{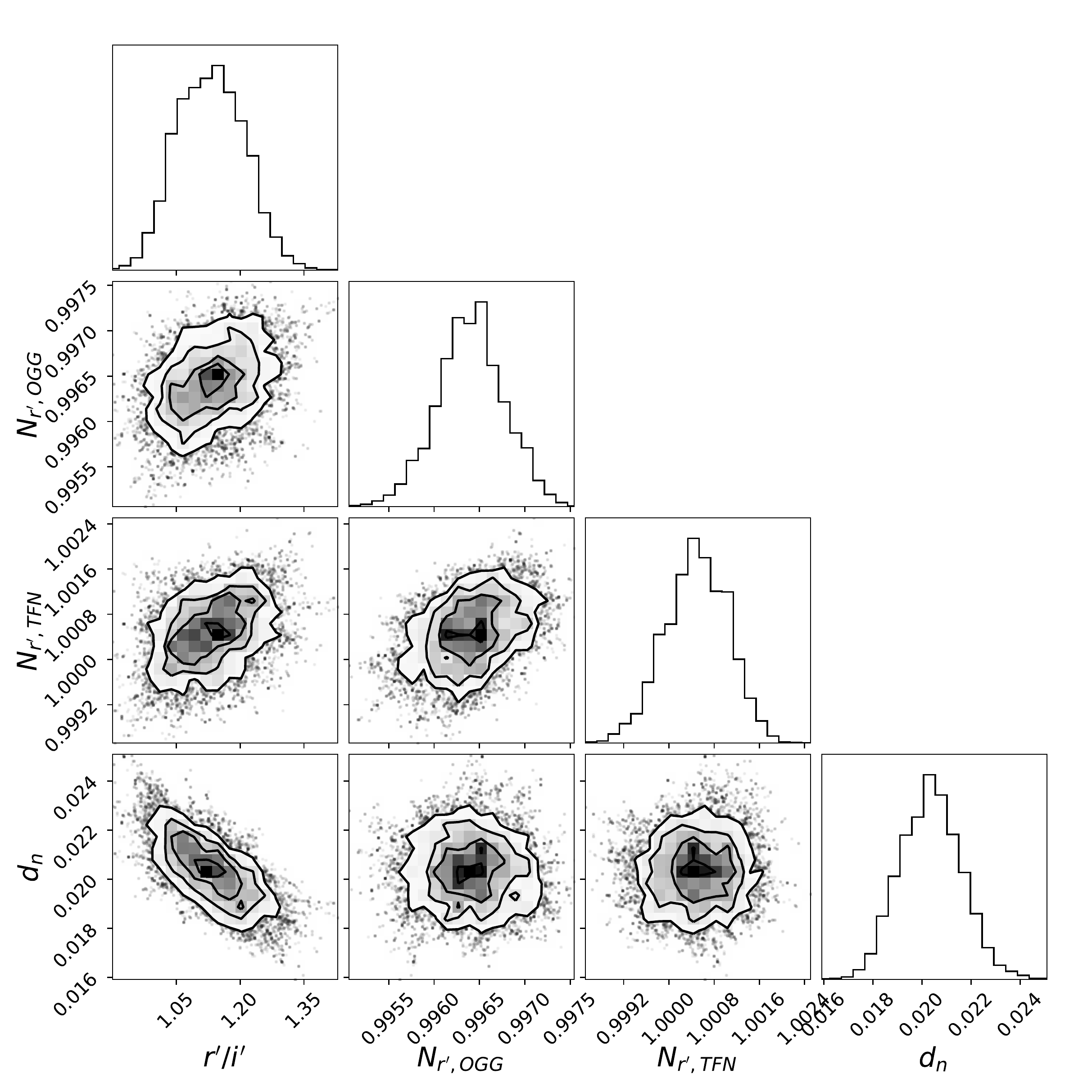}
\caption{Corner plot for the \angkor\ dip for the normalization in $r'$ band and $r'/i'$ color and the depth for the deepest day.
}
\label{fig:Arcorner}
\end{figure*}

\end{document}